# Joint MMSE Transceiver Design for Downlink Heterogeneous Network

Zhannan Li, Hangsong Yan, and I-Tai Lu

*Abstract*—In this paper, we propose a minimum mean square error (MMSE) based transceiver design scheme for a downlink multiple-input multiple-output (MIMO) two-tier heterogeneous network with general linear equality per-cell power constraints. Three practical channel models with both perfect and imperfect channel state information are used in simulations. In each channel model, we consider two system configurations, two data transmission schemes and two cellular cooperation scenarios. Our study shows that the proposed MMSE scheme is more flexible than interference alignment (IA) based scheme. For the cases where the IA-type scheme is applicable, the proposed scheme generally outperforms IA-type scheme in terms of average sum rate and bit error rate (BER), but is computationally more complex than the IA-type scheme.

*Index Terms*—Heterogeneous networks, MIMO, MMSE, interference alignment.

## 1 Introduction

A Heterogeneous network (HetNet) is a wireless network consisting of macrocells networks and some overlaid nodes with lower transmission powers [1]-[4]. It has emerged as a new trend in response to the explosive growth in data demands driven by smartphones, tablets, and various machine-type communication devices. Several studies have documented the astounding over 1000-fold increase in mobile data traffic demands in the last decades and have predicted that this trend will continue at a faster speed [5].

In a HetNet, the macrocell is deployed to provide a wide area coverage while the lower-tier small cells are deployed in a more targeted manner to alleviate call drops in blind zones and improve spectral efficiency of hot zones. Coordinated resource usage of macrocell, microcells, picocells and femtocells can significantly increase the channel capacity and extend the system coverage if the interference generated by the overlaid cells can be mitigated.

The interference in HetNet can be divided into two types: co-tier interference and cross-tier interference [6]. Co-tier interference represents interference occurring among cells in the same tier like picocells interfering with picocells. Cross-tier interference is the interference among cells in different tiers like macrocells interfering with picocells. Both of these two types of interference will significantly affect the spectral efficiency of a HetNet.

Zhannan Li (email: zhannan@hrbeu.edu.cn) is with Department of Information and Communication Engineering, Harbin Engineering University, Harbin, China.

Hangsong Yan (email: hy942@nyu.edu) and I-Tai Lu (email: itl211@nyu.edu) are with NYU WIRELESS Research Center, NYU Tandon School of Engineering, Brooklyn, NY 11201.

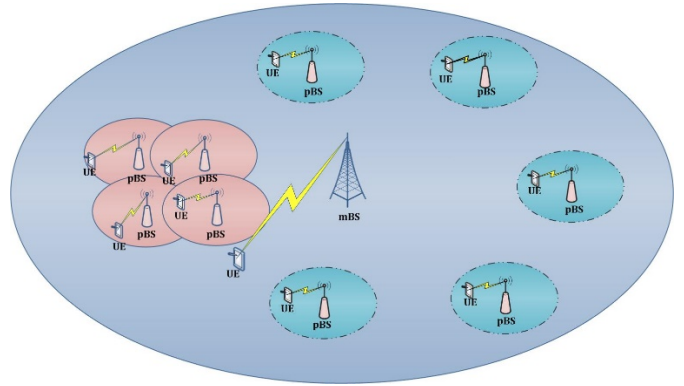

Fig. 1. A Heterogeneous Network Model

In recent years, interference alignment (IA) has attracted great attention for its application in managing the interferences in HetNet. For example, the work in [7]-[8] focused on applying IA to mitigate cross-tier interference. An IA based optimal transceiver design for cognitive radio (CR) networks was proposed in [9], which further increased the transmission rate of primary user and guaranteed its priority. References [10]-[11] divided small cells into some clusters and apply IA in each cluster to manage co-tier interference. A hierarchical IA (HIA) scheme was applied to address both the cross-tier and co-tier interference in HetNets [12]-[13]. However, these works are all limited to special configurations. Furthermore, IA is inherently quite restrictive [14]. When base stations (BSs) and user equipment (UEs) have limited numbers of equipped antennas, IA is infeasible for the entire system [15].

Alternatively, other types of transceiver designs in HetNets have been studied by many researchers. For example, in [16], a hierarchical precoding strategy was proposed to project interference from the macro BS (mBS) into the subspace of small cell users, which enabled linear cancellation. However, the proposed algorithms in [16] has high sensitivity to imperfect channel state information (CSI), which results in performance degradation in dense HetNets. A two-tier minimum mean square error (MMSE) precoding method was proposed in [17] to form deeper nulling for picocells close to mBS. However, the performance is based on the statistical CSI between mBS and pico BS (pBS). [18] considered open-loop spatial multiplexing (SM) with MMSE receiver for suppressing self-interference as well as other cell interference, but the data rate for sizable percentage of cell-edge UEs degraded when using full rate SM at mBS. In addition, second-order cone programming based and IA based transceiver designs were proposed in [19] for secure CR networks. Comparisons of the key features and performances between these two schemes were also presented. There is also research work about optimal transceiver designs in K-user multiple-input and multiple-output (MIMO) interference channels for simultaneous wireless information and power transfer (SWIPT) [20].

As IA and all of the above mentioned alternative approaches are somewhat limited, it would be beneficial to investigate *general* MMSE designs for HetNets. In this paper, we use the general iterative algorithm (GIA) in [21] to conduct various joint MMSE transceiver designs for two HetNet configurations. The first configuration is used in [15] which makes it possible to compare the GIA with the two stage IA (TSIA) in [15]. In the second configuration, the BSs and UEs are equipped with less antennas such that the TSIA becomes inapplicable. However, it will be shown that the GIA still yield good results. Note that the GIA is originally developed for homogeneous multicell networks. But it is extended here to various HetNet configurations. Both "with cooperation" and "without cooperation" scenarios are studied where the former can serve as a performance benchmark. Three channel models (uncorrelated, explicit correlation, and 3GPP) are used to investigate the average sum rate and bit error rate (BER) performances of GIA results. In each channel model, two different data transmission schemes, "partial" and "full", are considered for GIAs in both perfect and imperfect CSI conditions.

We have shown in this paper that the proposed MMSE approach has the following advantages over the IA-like approaches. Firstly, the proposed approach generally outperforms IA-like solutions in terms of average sum rate and BER, especially at low SNR regimes. Secondly, the proposed approach for different HetNet configurations can be formulated in the same way as an optimization problem and therefore a unified and systematic designing process can be used for different HetNet configurations. This is, however, not the case for IA-type designs (e.g., see [15]). As mentioned before, HetNets consist of macrocells and lower-tier small cells. The fact that some of the small cells clutter in different small areas for different HetNets make it very difficult to perform interference cancellation in the same way for various different HetNet configurations. Thirdly, the proposed approach has a less strict requirement on the antenna numbers at the BSs and UEs than the IA-type designs. For HetNet configurations where IA-type designs are not feasible due to insufficient degrees of freedom, the proposed approach may still yield satisfactory performances. Fourthly, the proposed approach allows for different numbers of data streams and are more flexible than the IA-type designs. Finally, the proposed approach allows for different cooperation scenarios and are more versatile than the IA-type designs. The only possible drawback of the proposed approach is that it is more computationally complex than the IA-type approaches.

Notations used in this paper are as follows: all boldface letters indicate vectors or matrices. $A', A^*, A^{-1}, tr(A)$, and $\langle A \rangle$ stand for the transpose, conjugate transpose, inverse, trace, and expectation of $A$, respectively. $null(A)$ denotes the nullspace of matrix $A$, $diag(...)$ represents the diagonal matrix with elements $(...)$ on the main diagonal. $I_a$ represents an identity matrix with rank $a$. $\lfloor a \rfloor$ is the largest integer no greater than $a$. $CN(\mu, \sigma^2)$ denotes a complex normal random variable with mean $\mu$ and variance $\sigma^2$.

## 2 Heterogeneous Network Model

### 2.1 Inter-Cell Interference in Heterogeneous Network

Shown in Fig.1 is an example of a two-tier HetNet [15] which consists of one macrocell and several overlaid picocells. Each cell has one BS and one UE. We assume a frequency division duplex (FDD) system. We also assume that all BSs use the same frequency band in the downlink and all UEs use the same frequency band in the uplink (which is different from the frequency band of the downlink). Each BS is located at the center of its corresponding cell; but each UE is randomly placed inside its corresponding cell.

Based on the co-tier interference (inter-cell interference among the picocells) for both downlink and uplink, the picocells are grouped into two weakly-coupled sub-systems. In sub-system 1, the picocells (plotted with solid lines) congregate in a small area and are likely to suffer strong inter-cell interferences from other picocells in the same sub-system. However, in sub-system 2, each picocell (plotted with dashed lines) is not close to any other picocell and will not suffer strong inter-cell interference from other picocells. This grouping of picocells into two sub-systems is needed for the TSIA approach but is not necessary for the MMSE approach.

Regarding the cross-tier interference (inter-cell interference between macrocell and picocells), downlink and uplink behave differently. In the downlink, the mBS will cause inter-cell interference to all pico UEs (pUEs) since all pUEs are located within the macrocell coverage range. On the other hand, the macro UE (mUE) may or may not suffer inter-cell interference from pBSs since its location is randomly generated and can be in any place inside the macrocell. In the uplink, the mBS will not suffer much inter-cell interference from pUEs as they are all far away from the mBS. However, some of the pBSs may suffer inter-cell interference from the mUE if the mUE happens to be located inside their cells.

### 2.2 Precoder and Decoder Formulation

For convenience and without loss of generality, we will study the example shown in Fig. 1 for the downlink scenario only. Let $s_i$ be the signal matrix intended for the $i^{th}$ UE in the $i^{th}$ cell where $i \in \{1, 2, ..., L\}$ and $L = L_1 + L_2$. Here, $i = 1$ denotes the macrocell, $i \in \{2, 3, ..., L_1\}$ denotes one of the $L_1 - 1$ picocells in sub-system 1, and $i \in \{L_1 + 1, L_1 + 2, ..., L\}$ denotes one of the $L_2$ picocells in sub-system 2. For the $i^{th}$ cell, the number of data streams is denoted by $m_i$. Then, $s_i \in \mathbb{C}^{m_i \times w}$ where $w$ is the number of data symbols of each data stream. The $m_i \times m_i$ source covariance matrix is defined as $\Phi_{si} = \langle s_i s_i^* \rangle$.

If there is full cooperation among all $L$ BSs, $s_i$ will be transmitted by all $L$ BSs [21]. If there is no full cooperation among the $L$ BSs, $s_i$ will be transmitted only by the $i^{th}$ BS. Then the received signal matrix at the $i^{th}$ UE is

$$y_i = \sum_{1 \leq j \leq L} h_{ij} \sum_{1 \leq k \leq L} f_{jk} s_k + n_i, \quad (1a)$$

in the with-cooperation scenario and

$$y_i = h_{ii}f_{ii}s_i + \sum_{j \neq i, 1 \leq j \leq L} h_{ij}f_{jj}s_j + n_i. \quad (1b)$$

in the without-cooperation scenario. Here, $n_i$ is the additive white Gaussian noise matrix with zero mean and its noise covariance matrix is $\Phi_{ni} = <n_i n_i^*> = \sigma_i^2 I_{r_i}$, where $\sigma_i^2$ is the noise variance. $h_{ij}$ is the channel matrix from BS $j$ to UE $i$. $f_{jk}$ is the transmit precoding matrix of BS $j$ for the signal matrix $s_k$. Let the number of receive antennas at the $i^{th}$ UE be denoted by $r_i$ and the number of transmit antennas at the $i^{th}$ BS be denoted by $t_i$. Then, both $y_i$ and $n_i \in \mathbb{C}^{r_i \times w}$, $f_{jk} \in \mathbb{C}^{t_j \times m_k}$, and $h_{ij} \in \mathbb{C}^{r_i \times t_j}$.

It is convenient to express both (1a) and (1b) in the following unified form:

$$y_i = H_i F_i s_i + \sum_{j \neq i, 1 \leq j \leq L} H_i F_j s_j + n_i. \quad (2)$$

where the channel matrix $H_i = [h_{i1} \ldots h_{i,i-1} \, h_{ii} \, h_{i,i+1} \ldots h_{iL}]$. The precoding matrix $F_j = [f_{1j}^* \ldots f_{j-1,j}^* \, f_{jj}^* \, f_{j+1,j}^* \ldots f_{Lj}^*]^*$ in the with-cooperation scenario. In the without-cooperation scenario, $F_j = [0_{1j}^* \ldots 0_{j-1,j}^* \, f_{jj}^* \, 0_{j+1,j}^* \ldots 0_{Lj}^*]^*$. Here, $0_{ij}^*$ is a zero matrix and is of the same dimension of $f_{ij}^*$. At the right hand side of the equal sign in (2), the first term is the desired signal and the second term is the sum of all inter-cell interference signals at UE $i$.

Let $G_i \in \mathbb{C}^{m_i \times r_i}$ be denoted as the decoder at UE $i$. Then the $m_i \times w$ decoded signal matrix is expressed as:

$$\tilde{y}_i = G_i y_i. \quad (3)$$

We are to find precoders $\{F_i\}$ and decoders $\{G_i\}$ for all cells with per cell total power constraint in order to satisfy either MMSE or IA criterion to mitigate inter-cell and intra-cell interferences. Sum rate $\sum_{i=1}^{L} C_i$ and BER will be computed as performance metrics to compare various MMSE designs against the IA design.

Here, $C_i$ is defined as

$$C_i = \log_2 \det(I_{m_i} + G_i H_i F_i \Phi_{si} F_i^* H_i^* G_i^* B^{-1}) \quad (4)$$

with $B = \sum_{j \neq i, 1 \leq j \leq L} G_i H_j F_j \Phi_{sj} F_j^* H_j^* G_i^* + \sigma_i^2 G_i G_i^*$. Note that $\hat{P}_j \equiv \text{tr}(F_j \Phi_{sj} F_j^*)$ denotes the total transmitting power used to transmit signal matrix $s_j$. In the without-cooperation scenario, $\hat{P}_j$ is reduced to $\text{tr}(f_{jj} \Phi_{sj} f_{jj}^*)$ and is equal to the total transmit power $P_j$ of BS $j$. Thus, for given $\{F_i\}$ and $\{G_i\}$, $C_i$ is the downlink capacity (i.e., the maximum achievable data rate) of cell $i$. But, in the with-cooperation scenario, all cells work together and each cell will transmit all $\{s_j\}$ with different precoders. The total power for transmitting $s_j$ from all cells, $\hat{P}_j$, which can also be expressed as $\sum_{i=1}^{L} \text{tr}(f_{ij} \Phi_{sj} f_{ij}^*)$, is different from the total transmit power $P_j \equiv \sum_{i=1}^{L} \text{tr}(f_{ji} \Phi_{si} f_{ji}^*)$ of BS $j$. Thus, for given $\{F_i\}$ and $\{G_i\}$, $C_i$ is not the downlink capacity of cell $i$, but is the maximum achievable data rate that can be carried by the signal matrix $s_i$.

### 2.3 CSI Channel Models

In practice, CSI is estimated and will result in CSI estimation error. In this work, we consider both perfect and imperfect CSI conditions. Let $h_{ij}$ and $\hat{h}_{ij}$, respectively, be the true and estimated channel matrices from the $j^{th}$ BS to the $i^{th}$ UE. Then,

$$h_{ij} = \hat{h}_{ij} + e_{ij}, \quad (5)$$

where $e_{ij}$ is the channel estimation error matrix.

We consider both correlated and uncorrelated channel models for channel matrices $\{h_{ij}\}$. For the correlated channel models, we further use both explicit correlation model and implicit correlation model where, in the former, the transmitting and receiving correlation matrices can be pre-specified.

#### 2.3.1 Explicit Correlation Model

To account for path loss and spatial correlation, the channel model with perfect CSI from the $j^{th}$ BS to the $i^{th}$ UE can be expressed as

$$h_{ij} = P_{L,ij} R_{R,i}^{1/2} h_{w,ij} R_{T,j}^{1/2}, \quad (6)$$

where $P_{L,ij}$ is the long term path loss. The matrix $R_{R,i}$ and $R_{T,j}$ are normalized (unit diagonal entries) transmit and receive correlation matrices. They are assumed to be full-rank. The entries of channel model $h_{w,ij}$ are i.i.d. $CN(0,1)$.

By using the well-established orthogonal training method, channel estimation error matrix is given as [22]

$$e_{ij} = P_{L,ij} R_{e,ij}^{1/2} e_{w,ij} R_{T,j}^{1/2}, \quad (7)$$

where $R_{e,ij}^{1/2} = [I_{r_i} + \sigma_{e,ij}^2 R_{R,i}^{-1}]^{-1}$ and the entries of $e_{w,ij}$ are i.i.d. $CN(0, \sigma_{e,ij}^2)$. Here, $\sigma_{e,ij}^2$ is the error variance. The estimated channel matrix $\hat{h}_{ij}$ is obtained from (5).

#### 2.3.2 Implicit Correlation Model

In addition to the explicit correlation model in (6), the ray-based 3GPP Spatial Channel Model (SCM) [23] is also used to generate the correlated channel matrix $h_{ij}$. The entries of the channel estimation error matrix $e_{ij}$ are i.i.d. $CN(0, \hat{\sigma}_{e,ij}^2)$. Similar to the explicit correlation model, the estimated channel matrix $\hat{h}_{ij}$ is obtained from (5). As the 3GPP SCM is based on the stochastic modeling of scatters [24], the correlations among the transmit antennas or the correlations among the receive antennas cannot be pre-specified but are implicitly determined by the statistical properties of ray parameters such as path powers, path delays, and angular characteristics. These ray parameters are modeled as random variables defined by their probability density functions (PDFs) and cross-correlations.

*2.3.3 Uncorrelated Model*

It is obvious that, by letting $R_{T,j}^{1/2} = I_{t_j}$ and $R_{R,i}^{1/2} = I_{r_i}$, the perfect uncorrelated channel model can be obtained from (6):

$$h_{ij} = P_{L,ij} h_{w,ij}, \qquad (8)$$

Similarly, the uncorrelated channel error matrix can be obtained from (7) by letting $R_{T,j} = I_{t_j}$ and $R_{e,ij}^{1/2} = I_{r_i}$:

$$e_{ij} = P_{L,ij} e_{w,ij} \qquad (9)$$

The estimated channel matrix $\hat{h}_{ij}$ is also obtained from (5).

## 3 MMSE Designs to Mitigate Inter-Cell Interference

### 3.1. MMSE Formulation with Imperfect CSI for HetNet

The MMSE designs are to choose precoders and decoders for minimizing the sum MSE $\eta$ of the HetNet

$$\{G_{i,mmse}, F_{i,mmse}\} = \underset{\{G_i, F_i\}}{\operatorname{argmin}}(\eta) \qquad (10)$$

subjected to the per-cell power constraint

$$P_j = \sum_{i=1}^{L} \operatorname{tr}(f_{ji} \Phi_{si} f_{ji}^*) \text{ or } P_j = \operatorname{tr}(f_{jj} \Phi_{sj} f_{jj}^*) \qquad (11)$$

of BS $j$, $\forall j \in \{1, ..., L\}$, for "with cooperation" and "without cooperation" scenarios, respectively. For reasons to be clear later, it is more conveniently to express the power constraints for both scenarios in (11) in the following unified matrix form:

$$\operatorname{tr}(Q_j P Q_j) = \operatorname{tr}\left(Q_j \left(\sum_{i=1}^{L} F_i \Phi_{si} F_i^*\right) Q_j\right) \qquad (12)$$

where the power matrix $P = \operatorname{diag}(p_1 ... p_{j-1} p_j p_{j+1} ... p_L)$ with $p_j = I_{t_j} P_j / t_j$ $\forall j \in \{1, ..., L\}$, and the diagonal selection matrix $Q_j = \operatorname{diag}(\tilde{q}_1 ... \tilde{q}_{j-1} q_j \tilde{q}_{j+1} ... \tilde{q}_L)$ with $q_j = I_{t_j}$ and $\tilde{q}_j = 0_{t_j}$. Note that the multiplications of $Q_j$ from both sides of a matrix is to select the $j^{th}$ diagonal block of that matrix.

In (10) the sum MSE $\eta$ is defined as

$$\eta = \sum_{1 \leq i \leq L} \eta_i \qquad (13)$$

and $\eta_i$ is the MSE of the $i^{th}$ cell

$$\eta_i = \operatorname{tr}\langle(G_i y_i - s_i)(G_i y_i - s_i)^*\rangle. \qquad (14)$$

Here, $y_i$ is the received signal matrix at the $i^{th}$ UE while the data streams $s_i$ is intended for the $i^{th}$ UE as defined in (2). When CSI is imperfect, it is more convenient to express $y_i$ in (2) in the following alternative form for the MMSE design to be presented later:

$$y_i = \hat{H}_i F_i s_i + \sum_{j \neq i, 1 \leq j \leq L} \hat{H}_i F_j s_j + \dot{n}_i, \qquad (15)$$

where the estimated channel matrix $\hat{H}_i = [\hat{h}_{i1} ... \hat{h}_{iL}]$. The second term in the right hand side of (15) is the equivalent interference. The last term in (15) is the equivalent noise matrix and is defined as

$$\dot{n}_i = n_i + E_i \sum_{1 \leq j \leq L} F_j s_j \qquad (16)$$

where the channel estimation error matrix $E_i = [e_{i1} ... e_{iL}]$. After some mathematical manipulations, equation (13) becomes

$$\eta_i = \operatorname{tr}\langle \Phi_{si} - G_i \hat{H}_i F_i \Phi_{si} - \Phi_{si} F_i^* \hat{H}_i^* G_i^* \rangle$$
$$+ \operatorname{tr}\langle G_i \left( \hat{H}_i \left( \sum_{j \in S} F_j \Phi_{sj} F_j^* \right) \hat{H}_i^* + \dot{\Phi}_{ni} \right) G_i^* \rangle \qquad (17)$$

Where $S = \{j\}$, $\dot{\Phi}_{ni} = \langle \dot{n}_i \dot{n}_i^* \rangle + \langle \hat{H}_i (\sum_{j \neq i, 1 \leq j \leq L} F_j \Phi_{sj} F_j^*) \hat{H}_i^* \rangle$ for "without cooperation" case, and $S = \{1, ..., L\}$, $\dot{\Phi}_{ni} = \langle \dot{n}_i \dot{n}_i^* \rangle$ for "with cooperation" case.

### 3.2 MMSE Precoder and Decoder Design

Setting the gradient of $\eta$ in (13) with respect to $G_i$ equal to zero, for the given set of precoder $\{F_j\}$, we can obtain the MMSE decoder as

$$G_i = \Phi_{si} F_i^* \hat{H}_i^* M_i,$$
$$M_i = \left( \hat{H}_i \left( \sum_{j \in S} F_j \Phi_{sj} F_j^* \right) \hat{H}_i^* + \dot{\Phi}_{ni} \right)^{-1}. \qquad (18)$$

Note that $G_i = G_{i,mmse}$ in (18) only if $\{F_i = F_{i,mmse}\}$. Substitute (18) into (13) we get a cost function which does not depend on $G_i$ as

$$\hat{\eta} = \sum_{j \in L} \operatorname{tr}(\Phi_{sj} - \Phi_{sj} F_j^* \hat{H}_j^* M_j \hat{H}_j F_j \Phi_{sj}). \qquad (19)$$

Using (19), one can use the method of Lagrange multipliers to formulate an augmented cost function for solving (10) and (12) as below:

$$\{F_{i,mmse}\} = \underset{\{F_i\}}{\operatorname{argmin}}(\hat{\xi}) \qquad (20)$$

where

$$\hat{\xi} = \hat{\eta} + \operatorname{tr}\left( \Lambda \left( \sum_{1 \leq j \leq L} F_j \Phi_{sj} F_j^* - P \right) \right). \qquad (21)$$

Here $\Lambda = \operatorname{diag}(\lambda_1 ... \lambda_L)$ where $\lambda_j = I_{t_j} \lambda_j$ and $\{\lambda_j\}$ are Lagrange multipliers. We can then set the gradient of $\hat{\xi}$ in (20) with respect to $F_i$ equal to zero to get $L$ equations of $\{F_{i,mmse}\}$ and $\{\lambda_j\}$. Together with the $L$ power constraints in (12), one can in principle solve the $L$ optimum precoders $\{F_{i,mmse}\}$. However, this is a set of $2L$ coupled nonlinear equations and is very difficult to solve. Instead, we will use the following iterative approach to solve for $\{G_{i,mmse}\}$ and $\{F_{i,mmse}\}$.

### 3.3 Alternative Iterative Approach

Instead of using (19-21), we use the method of Lagrange

multipliers to formulate an alternative augmented cost function for solving (10) and (12) as below:

$$\{G_{i,mmse}, F_{i,mmse}\} = \underset{\{G_i, F_i\}}{\text{argmin}}(\xi) \quad (22)$$

where

$$\xi = \eta + tr\left(\Lambda\left(\sum_{1 \le j \le L} F_j \Phi_{sj} F_j^* - P\right)\right) \quad (23)$$

Here, $\eta$ is given in (13), $\Lambda$ is given in (21), and $P$ is given in (12). The MMSE precoders can then be obtained by setting the gradient of $\xi$ in (23) with respect to $F_i$ equal to zero for the given sets of decoders $\{G_i\}$ and Lagrange multipliers $\{\lambda_j\}$. Thus, we have

$$F_i = N\widehat{H}_i^* G_i^*, \quad N = \left(\sum_{j \in S} \widehat{H}_j^* G_j^* G_j \widehat{H}_j + \Lambda\right)^{-1}. \quad (24)$$

In addition, we set the gradient of $\hat{\xi}$ in (21) with respect to $F_j$ equal to zero, left multiplying $F_j$ to the resulting equation, and summing up the resulting equation over $j$, we obtain

$$\sum_{1 \le j \le L} (F_j \Phi_{sj} F_j^*) \Lambda = D, \quad (25)$$

where

$$D = \sum_{j \in S} F_j \Phi_{sj}^2 F_j^* \widehat{H}_j^* M_j \widehat{H}_j - \left(\sum_{j \in S} F_j \Phi_{sj} F_j^*\right) \cdot \sum_{l \in S} \widehat{H}_l^* M_l \widehat{H}_l F_l \Phi_{sl}^2 F_l^* \widehat{H}_l^* M_l \widehat{H}_l$$

Based on equation (25) and the definition of power constraint defined in (12), we can obtain explicit expressions for the Lagrange multipliers $\lambda_j$ with respect to the per-cell power constraint as

$$\lambda_j = P_j^{-1} tr[Q_j D Q_j], \quad (27)$$

where the multiplications of the selection matrix $Q_j$ from both sides of (12) is to select the $j^{th}$ diagonal block of $D$. The generalized iterative approach (GIA) uses the expressions in (18), (24) and (27) to search for sub-optimum precoders and decoders jointly.

There are three steps in each iteration of the *GIA*.
 *Step 1.* Given $\{F_j\}_{1 \le j \le L}$, obtain $\{G_j\}_{1 \le j \le L}$ by (18).
 *Step 2.* Given $\{F_j\}_{1 \le j \le L}$, obtain $\{\lambda_j\}$ or $\Lambda$ by (27).
 *Step 3.* Given $\{G_j\}_{1 \le j \le L}$ and $\Lambda$, obtain $\{F_j\}_{1 \le j \le L}$ by (24)

## 4 IA Designs to Mitigate Inter-Cell Interference

Note that there is no cooperation in the IA scheme. That is the signal matrix $s_j$ is transmitted by BS $j$ only. To achieve IA for the HetNet in Fig. 1, we need to satisfy the following four equations:

$$G_k h_{kj} f_{jj} = 0, \quad k \ne j, 1 \le k, j \le L_1 \quad (28)$$
$$G_k h_{k1} f_{11} = 0, \quad L_1 + 1 \le k \le L \quad (29)$$
$$G_1 h_{1j} f_{jj} = 0, \quad L_1 + 1 \le j \le L \quad (30)$$
$$rank(G_j h_{jj} f_{jj}) = m_j, \quad 1 \le j \le L. \quad (31)$$

Here, (28) mitigates the mutual inter-cell interference of the picocells in sub-system 1 and the inter-cell interference between the macrocell and the picocells in sub-system 1, (29) mitigates the interference from the mBS to the pUEs in sub-system 2, (30) mitigates the interference from the pBSs in sub-system 2 to the mUE, and (31) guarantees that the signal space $G_j h_{jj} f_{jj}$ has $m_j$ dimensions and is orthogonal to the corresponding interference subspace. In practice, there are usually not enough transmit and/or receive antennas to satisfy all of these conditions. Thus, we can only use (28)-(31) as design guidelines for minimizing the inter-cell interferences of the HetNet system.

Following the two stage interference alignment (TSIA) approach in [15], in the first stage, some degrees of freedom in mBS are leveraged to nullify the interference from mBS to as many pUEs in sub-system 2 as possible. Using (29), let $f_{11} = V_1 \hat{f}_{11}$, where the null space to pUEs $\{L_1+1, L_1+2, \ldots, L_1+n\}$ is

$$V_1 = null\left(\left[h'_{L_1+1,1}, \ldots, h'_{L_1+n,1}\right]'\right), \quad (32)$$

where $n$ is the maximum number of pUEs in sub-system 2 with interference from mBS being canceled. Here, we define

$$n = min\left\{\left|\frac{t_1 - t_j}{r_j}\right|_{L_1+1 \le j \le L}, L_2\right\}. \quad (33)$$

Then, the equivalent channel matrices from mBS to all UEs is $\tilde{h}_{j1} = h_{j1} V_1, 1 \le j \le L$, and the equivalent precoder $\tilde{f}_{11}$ is to be determined.

Furthermore, the precoders and decoders of macrocell and all picocells in sub-system 1 are designed to satisfy (28). By using the conventional IA algorithms [25], the precoders and decoders are iteratively updated by minimizing the total interference leakages $E_i$ at UE $i$ in the downlink and $\overleftarrow{E}_j$ at BS $j$ in the reverse link with $1 \le i, j \le L_1$,

$$E_i = tr(G_i Z_i G_i^*), \quad \overleftarrow{E}_j = tr(\overleftarrow{G}_j \overleftarrow{Z}_j \overleftarrow{G}_j^*)$$

$$Z_1 = \sum_{j=2}^{L_1} P_j h_{1j} f_{jj} f_{jj}^* h_{1j}^*$$

$$Z_{i(i \ne 1)} = P_1 \tilde{h}_{i1} \tilde{f}_{11} \tilde{f}_{11}^* \tilde{h}_{i1}^* + \sum_{j=2, j \ne i}^{L_1} P_j h_{ij} f_{jj} f_{jj}^* h_{ij}^*$$

$$\overleftarrow{Z}_j = \sum_{i=1, i \ne j}^{L_1} P_i \overleftarrow{h}_{ji} \overleftarrow{f}_{ii} \overleftarrow{f}_{ii}^* \overleftarrow{h}_{ji}^* \quad (34)$$

with $\overleftarrow{G}_j = f_{jj}^*, \overleftarrow{h}_{ij} = h_{ji}^*, \overleftarrow{f}_{jj} = G_j^*$ for $2 \le j \le L_1$ and $1 \le i \le L_1$. Also $\overleftarrow{G}_1 = \tilde{f}_{11}^*, \overleftarrow{h}_{1i} = \tilde{h}_{i1}^*, \overleftarrow{f}_{11} = G_1^*$ for the macrocell. To minimize $E_i$ and $\overleftarrow{E}_j$, we choose $G_i = v_d(Z_i)$ and $\overleftarrow{G}_j = v_d(\overleftarrow{Z}_j)$, and $v_d(A)$ is the matrix formed by the $m_i$ eigenvectors corresponding to the $m_i$ smallest eigenvalues of $\boldsymbol{A}$.

In the second stage, we are to design precoders and decoders for sub-system 2. At first, based on (30), pBSs should align their transmitting signals to the interference subspace $\boldsymbol{U}_j$ of mUE in order to cancel the interference from pBSs in sub-system 2 to the mUE:

$$\boldsymbol{U}_j = null(\boldsymbol{G}_1 \boldsymbol{h}_{1j}), \quad L_1 + 1 \leq j \leq L. \quad (35)$$

Thus, the precoders of pBSs in sub-system 2 can be expressed as

$$\boldsymbol{f}_{jj} = \boldsymbol{U}_j \tilde{\boldsymbol{f}}_{jj}, \quad L_1 + 1 \leq j \leq L. \quad (36)$$

Recall that, for $L_1 + 1 \leq j \leq L_1 + n$, the interference from mBS to pUEs in these picocells are nullified. Thus, various joint precoder and decoder designs can be adopted for designing $\tilde{\boldsymbol{f}}_{jj}$ and $\boldsymbol{G}_j$. For example, define the equivalent channel matrices as

$$\tilde{\boldsymbol{h}}_{jj} = \boldsymbol{h}_{jj} \boldsymbol{U}_j, \quad L_1 + 1 \leq j \leq L_1 + n. \quad (37)$$

Apply singular value decomposition (SVD) on $\tilde{\boldsymbol{h}}_{j,j}$ and obtain

$$\tilde{\boldsymbol{h}}_{jj} = \tilde{\boldsymbol{G}}_j \boldsymbol{\varepsilon}_j \tilde{\boldsymbol{f}}_{jj}^*, \quad L_1 + 1 \leq j \leq L_1 + n. \quad (38)$$

Therefore, the SVD precoder and decoder can be expressed as $\boldsymbol{f}_{jj} = \boldsymbol{U}_j \tilde{\boldsymbol{f}}_{jj}$ and $\boldsymbol{G}_j = \tilde{\boldsymbol{G}}_j$.

As for $L_1 + n \leq j \leq L$, the interference from mBS to pUEs in these picocells are *not* nullified. Thus, the pUEs interfered by mBS have to cancel the interference via their decoders. Therefore, based on (29), the desired signal should be received in the orthogonal complement space of macrocell signal space, that is

$$\boldsymbol{T}_j = null\big((\boldsymbol{h}_{j1} \boldsymbol{f}_{11})^*\big), \quad L_1 + n < j \leq L. \quad (39)$$

Thus, decoders of pUEs interfered by mBS is expressed as

$$\boldsymbol{G}_j = \boldsymbol{T}_j \tilde{\boldsymbol{G}}_j, \quad L_1 + n < j \leq L. \quad (40)$$

Define the equivalent channel matrix $\tilde{\boldsymbol{h}}_{jj}$ as

$$\tilde{\boldsymbol{h}}_{jj} = \boldsymbol{T}_j^* \boldsymbol{h}_{jj} \boldsymbol{U}_j, \quad L_1 + n < j \leq L. \quad (41)$$

Perform SVD on the equivalent channel matrix $\tilde{\boldsymbol{h}}_{jj}$

$$\tilde{\boldsymbol{h}}_{jj} = \tilde{\boldsymbol{G}}_j \boldsymbol{\varepsilon}_j \tilde{\boldsymbol{f}}_{jj}^*, \quad L_1 + n < j \leq L. \quad (42)$$

Thus, precoders and decoders of picocells interfered by mBS can be expressed as $\boldsymbol{f}_{jj} = \boldsymbol{U}_j \tilde{\boldsymbol{f}}_{jj}$ and $\boldsymbol{G}_j = \boldsymbol{T}_j \tilde{\boldsymbol{G}}_j$, respectively.

Finally, the power constraints are satisfied by choosing a diagonal matrix $\boldsymbol{\Phi}_{sj}$ such that $P_j = \text{tr}(\boldsymbol{f}_{jj} \boldsymbol{\Phi}_{sj} \boldsymbol{f}_{jj}^*)$. Since $\boldsymbol{f}_{jj}$ is unitary or semi-unitary in sub-system 2, the water-filling power allocation scheme over singular value matrix $\boldsymbol{\varepsilon}_j$ in (38) or (42) can be used to improve the achievable data rate.

## 5 Numerical Results

Consider a HetNet consisting of nine picocells (with cell indexes $i$ = 2, …, 10) and one macrocell (with cell index $i$ = 1) as shown in Fig. 1. The radius of the macrocell is 500 m and the radius of the picocell is 40 m. There is a hotspot area consisting of four picocells (with cell indexes $i$ = 2, …, 5) and the radius of the hotspot area is 100 m. Note that mUE is randomly distributed in the hotpot area in our simulation. Distance between the mBS and the hotspot center is 350 m. These closely arranged four picocells make up sub-system 1. In addition, there are five picocells (with cell indexes $i$ = 6, …,10) in sub-system 2; they are distributed in a circle with a radius of 350 m as shown in Fig. 1. Each BS serves one UE which is randomly distributed in the corresponding coverage area of each cell. Thus, $L_1 = L_2 = 5$ and $L = 10$ in this example.

The per-cell transmission power range for picocells is between 0 to 40 dBm. To reflect the performances of GIA and TSIA more intuitively, we also use the effective transmit SNR (transmit power in dBm + path loss in dB – noise power in dBm) as the reference for performance evaluation.

For both perfect and imperfect CSI conditions, we compare the average sum rate and BER performances of GIA and TSIA whenever TSIA is feasible. In our simulation, identical path loss and noise models for the three channel models as discussed in Section 2.3 are used. In this way, the effective transmit SNRs are the same for the three channel models and the performances of proposed algorithms in different channel models can be compared directly. In addition, the variances $\sigma_{e,i,j}^2$ and $\tilde{\sigma}_{e,i,j}^2$ of channel estimation errors in Section 2.3 are set as $10^{-3}$ throughout the transmission power range. Detailed simulation parameters are provided in Table 1.

Table 1
SIMULATION PARAMETERS

| Parameter | Assumption |
| --- | --- |
| Radius of macrocell | 500 m |
| Radius of picocell | 40 m |
| Carrier frequency/bandwidth | 2 GHz/100 MHz |
| Path loss from macro BS to user | 128.1 + 37.6log$_{10}$R[dB], R in km |
| Path loss from pico BS to user | 140.7 + 36.7log$_{10}$R[dB], R in km |
| Antenna pattern | 0 dB (omni-directional) |
| Shadowing standard deviation | 10 dB |
| Noise spectral density | -174 dBm/Hz |

### 5.1 "Sufficient" Configuration

In our simulation, we consider a MIMO configuration with sufficient numbers of transmit and receive antennas in each cell for performing the TSIA design. Following an example in [15], the number of transmit antennas at the mBS $t_1 = 6$, the number of transmit antennas at each pBS $t_i = 3$, for $i = 2, …, 10$, and the number of receive antennas at each UE $r_j = 3$, for $j = 1, …, 10$. Two data transmission schemes are considered for the GIA approach. The first data transmission scheme is defined as follows: $m_i = 1$ with $i \in \{1,2,…,5\}$ and $m_i = 2$ with $i \in \{6,7,…,10\}$. The second data transmission scheme is defined as follows: $m_i = r_i = 3$ for $i \in \{1,…,10\}$. Since the TSIA approach is less flexible than the GIA approach, only the first data transmission scheme is considered for the TSIA approach. This is due to the fact that the TSIA approach cannot handle the second data transmission scheme with the limited numbers of antennas given above.

Under the perfect CSI condition where there is no channel estimation error, there will be four GIA designs since we can have "with cooperation" and "without cooperation" two scenarios as formulated in (1a) and (1b); furthermore, each of the two scenarios can have two data transmission schemes as mentioned above. Here, the first data transmission scheme will be denoted as "partial" since the number of data streams transmitted is less than the degree of freedom in each cell (i.e., $m_i < \min(t_i, r_i)$); and the second data transmission scheme will be denoted as "full" since the number of data streams transmitted is equal to the degree of freedom in each cell (i.e., $m_i = \min(t_i, r_i)$). Thus, the four GIA designs will be denoted as: GIA with cooperation full, GIA with cooperation partial, GIA without cooperation full, and GIA without cooperation partial. Similarly, under imperfect CSI condition where there is CSI estimation error, there will also be four GIA designs: NGIA with cooperation full, NGIA with cooperation partial, NGIA without cooperation full, and NGIA without cooperation partial. As for the TSIA, only the TSIA without cooperation partial is available. It will be simply denoted as TSIA under perfect CSI condition and NTSIA under imperfect CSI condition. Here, "N" denotes naive since we do not try to compensate for or mitigate the channel estimation error here.

*5.1.1 Uncorrelated channel model*

Fig. 2 shows the average sum rates of GIA and TSIA under different effective transmit SNRs of picocells. Firstly, the sum rates of four "with cooperation" curves are larger than the sum rates of the rest of six "without cooperation" curves (including the TSIA and NTSIA) at high effective transmit SNRs. It shows that MMSE-type "with cooperation" algorithms can mitigate the interference better than the IA-type algorithms or the MMSE-type "without cooperation" algorithms. As the effective transmit SNR increases, the "with cooperation" sum rates continue to grow (in the given parameter regimes) but the "without cooperation" sum rates become saturated at high effective transmit SNRs. This means that the "with cooperation" algorithms are noise-limited, but the "without cooperation" are interference-limited. This is due to the fact that under the "with cooperation" scenario, the entire HetNet becomes a large single-user MIMO system. Therefore, there is no more inter-cell interference, only the inter-data-stream interference remains to be dealt with.

Secondly, comparing the MMSE results of the same category but with different data transmission schemes, the MMSE with "full" data transmission scheme achieves a higher average sum rate than the MMSE with "partial" data transmission scheme. It is because the "full" data transmission scheme fully utilizes the degree of freedom provided by the HetNet MIMO system.

Thirdly, due to the full utilization of the degree of freedom, the sum rates of MMSE without cooperation full algorithm is better than the results of IA and MMSE without cooperation partial throughout the whole effective transmit SNR range.

Fourthly, the performances of MMSE without cooperation partial are similar to IA. MMSE results are better than the IA at low transmit power where the noise effect is dominant and the system is noise-limited. IA is better than MMSE at high SNRs because the TSIA has taken the hot spot in Fig. 1 into consideration but the MMSE treats every cell equally. If we also take the hot spot into consideration by weighting the MSEs differently for different UEs in the MMSE design, MMSE result can be no worse than IA for any SNR.

Finally, comparing the same type of algorithm under different CSI conditions, it is obvious that the perfect CSI result is better than the imperfect CSI result. Since the interference power increases with the transmit power and the rate of increase of interference is larger for the imperfect CSI condition than for the perfect CSI condition for the same type of algorithm, the difference of sum rates between perfect and imperfect CSI conditions for the same type of algorithm increases as the transmit power increases. This phenomenon is profound for MMSE "with cooperation", but is almost invisible for MMSE "without cooperation". For the former, this is because the system is noise-limited in the entire SNR range under perfect CSI condition, but is interference-limited for large SNRs under the imperfect CSI condition. For the latter, the system is primarily interference-limited in the entire SNR range for both perfect and imperfect CSI conditions, and the CSI error is very small in our simulation. Therefore, no more interference is significantly created at large SNRs.

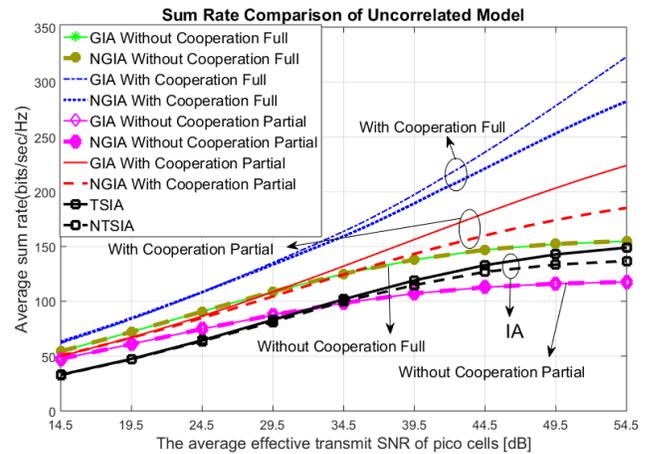

Fig. 2. Average sum rates vs. the effective transmit SNR of picocells in uncorrelated channel model.

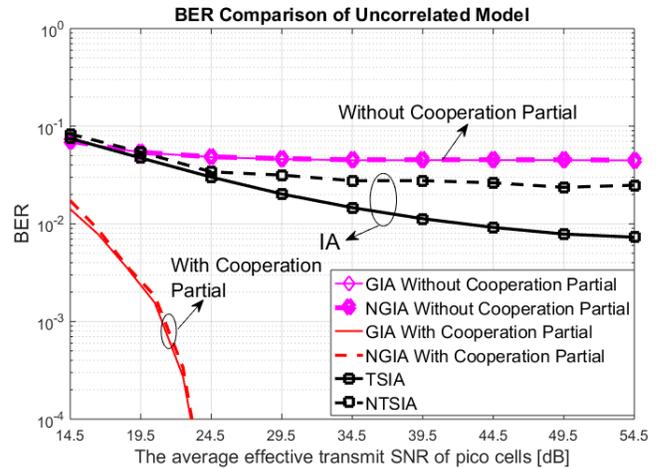

Fig. 3. BER vs. the effective transmit SNR of picocells in uncorrelated channel model.

Fig. 3 shows the raw BER performances of GIA "partial" and TSIA under different effective transmit SNRs of picocells since they have the same number of data streams and therefore the same number of total data bits. Firstly, the BER performances of "with cooperation" are much better than those of "without cooperation" which include the TSIA and NTSIA. As effective transmit SNR increases, the BERs of "with cooperation" drop sharply, but the BERs of "without cooperation" hit the error floors quickly. Again, this means that the "with cooperation" algorithms are noise-limited, but the "without cooperation" are interference-limited.

Secondly, comparing the GIA "without cooperation" and TSIA, the GIA works better in low effective transmit SNRs and the TSIA works better in high effective transmit SNRs. Again, this shows the MMSE "without cooperation" deals with the combined effect of interference and noise while the IA deals with interference only. As mentioned before, the TSIA has taken the hot spot in Fig. 1 into consideration but the MMSE treats every cell equally. If we also take the hot spot into consideration by weighting the MSEs differently for different UEs in the MMSE design, MMSE result can be no worse than IA for any SNR (e.g., see [14]).

Note that the raw BER performances of GIA "full" are not shown here because the raw BERs of the weak eigen channels will overshadow the raw BERs of the strong eigen channels in a MIMO system with multiple data streams. In this case, only the BERs of coded systems with appropriate modulation coding schemes (MCSs) can properly show the performance of the system.

### 5.1.2 Explicit correlation channel model

Let the receive and transmit correlation matrices $\boldsymbol{R}_{R,i}$ and $\boldsymbol{R}_{T,j}$ in (6) and (7) be normalized Toeplitz matrices with correlation coefficient being 0.6. Fig. 4 shows the average sum rates of GIA and TSIA, and Fig. 5 shows raw BER performances of GIA "partial" and TSIA under different effective transmit SNRs of picocells. The observations made in previous sub-section for the uncorrelated channel model also hold for the explicit correlation channel model and will not be repeated here. Only two additional comments on the effects of correlated channels will be made below. Firstly, comparing Fig. 4 against Fig. 2 for the same type of algorithms, the sum rates of correlated channels in Fig. 4 are usually lower than the sum rates of uncorrelated channels in Fig. 2. This is due to the fact that the condition number of a correlated channel matrix (average: 11.8 dB for picocell or 7.0 dB for macrocell) is usually larger than the condition number of an uncorrelated channel matrix (average: 9.0 dB for picocell or 4.1 dB for macrocell). Secondly, the sum rate differences between MMSE with cooperation "full" and "partial" in Fig. 4 are reduced compared with the results in Fig. 2. When effective transmit SNR of picocells equals 54.5 dB, the sum rate differences between MMSE with cooperation "full" and "partial" are 98.97 bits/sec/Hz and 66.51 bits/sec/Hz in Fig. 2 and Fig. 4, respectively. This phenomenon can also be explained by larger condition number of explicit correlation model compared with uncorrelated model. When the condition number of a MIMO channel matrix is large, the channel gains of the weaker eigen channels (i.e., equivalent Single Input Single Output (SISO) channels to original MIMO channel matrix) are small. This means the total sum capacity will be small and the gap between "full" and "partial" will be small too.

### 5.1.3 Implicit Correlation Model (3GPP Model)

In this section, we investigate the average sum rate and BER performances of GIA and TSIA in the 3GPP channel model [23]. Fig. 6 and Fig. 7 show the average sum rates of GIA and TSIA, and raw BER performances of GIA "partial" and TSIA under different effective transmit SNRs of picocells, respectively. Similar observations can also be made in the 3GPP model as in the uncorrelated and explicit correlation models, and will not be repeated here. There is one exception that the sum rates of MMSE without cooperation full is reduced to be below the TSIA curve at high SNRs. As explained previously, this is due to the fact that TSIA takes advantage of the distribution of the hot spot in the transceiver design. If the MMSE design also takes the hot spot into consideration by means of weighting the MSEs differently for different UEs, IA will not outperform the MMSE design for any SNR. In addition to this, when comparing Fig. 6 against Fig. 4 and Fig. 2 for the same type of algorithms, the sum rates of the 3GPP correlated channels in Fig. 6 are usually lower than the sum rates of the explicit correlation channels in Fig. 4 and the uncorrelated channels in Fig. 2 at the same effective transmit SNR. This is because the condition numbers of channel matrices for the 3GPP models (average: 21.2 dB for picocell or 13.5 dB for macrocell) are generally higher than those in the explicit correlation and uncorrelated channel matrices in the previous sections. Moreover, due to the large condition number of 3GPP channels, the sum rate differences between MMSE with cooperation "full" and "partial" is further reduced in Fig. 6. The difference is 39.18 bits/sec/Hz at 54.5 dB effective transmit SNR of picocells, which is much smaller than the corresponding results of explicit correlation and uncorrelated channels shown in the previous section and in Figs. 2 and 4.

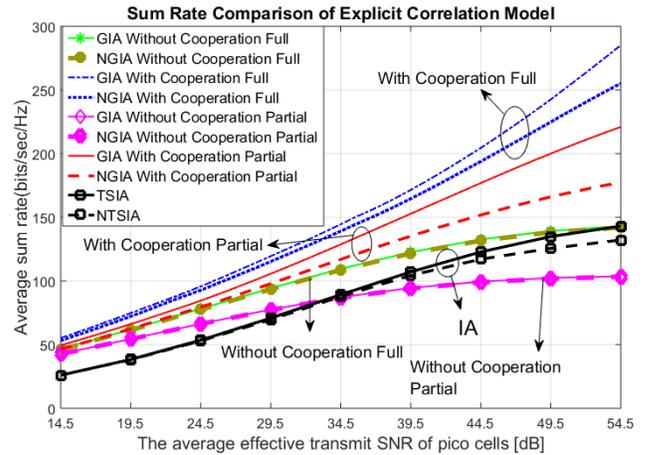

Fig. 4. Average sum rates vs. the effective transmit SNR of picocells in explicit correlation channel model.

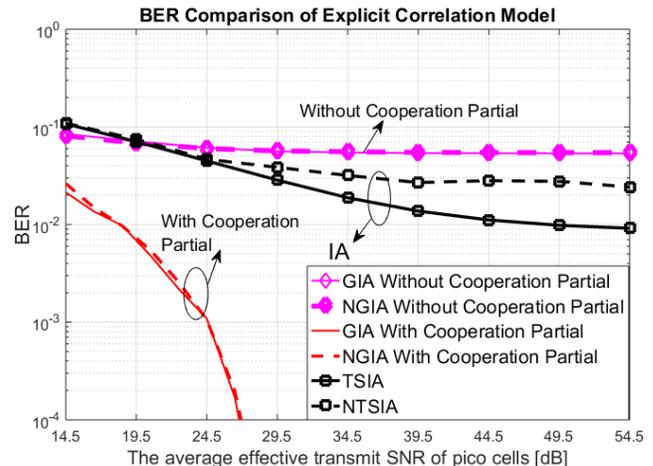

Fig. 5. BER vs. the effective transmit SNR of picocells in explicit correlation channel model.

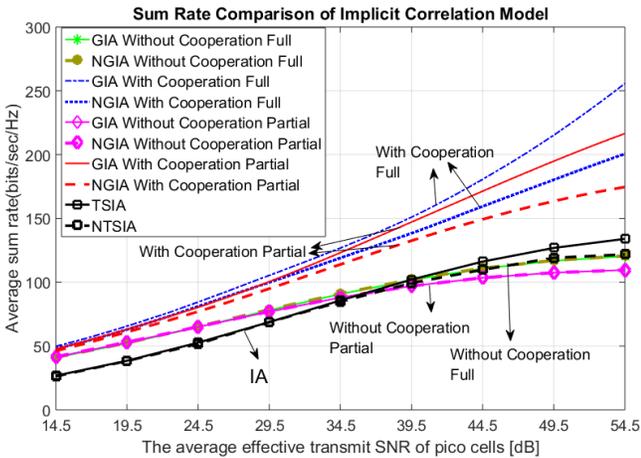

Fig. 6. Average sum rates vs. the effective transmit SNR of picocells in implicit correlation channel model.

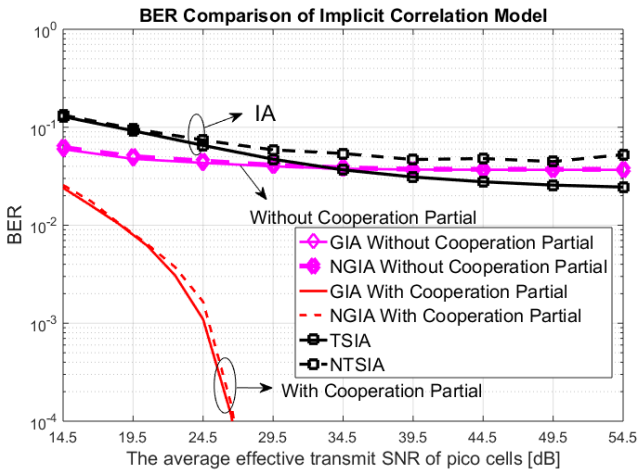

Fig. 7. BER vs. the effective transmit SNR of picocells in implicit correlation channel model.

## 5.2 "Insufficient" Configuration

Consider a MIMO configuration without sufficient numbers of transmit and receive antennas in each cell for performing the TSIA design. Modifying the example in the previous section, the number of transmit antennas at mBS $t_1 = 4$, the number of transmit antennas at pBSs $t_i = 2$ for $i = 2, \ldots, 10$, and the number of receive antennas at both mUE and pUEs $r_j = 2$ for $j = 1, \ldots, 10$. Two data transmission schemes are considered for the GIA approach. The first data transmission scheme denoted as "partial" is defined as follows: $m_i = 1$ with $i \in \{1,2,\ldots,5\}$ and $m_i = 2$ with $i \in \{6, \ldots, 10\}$. The second data transmission scheme denoted as "full" is defined as follows: $m_i = r_i = 2$ for $i \in \{1, ..., 10\}$. Note that TSIA is no longer feasible.

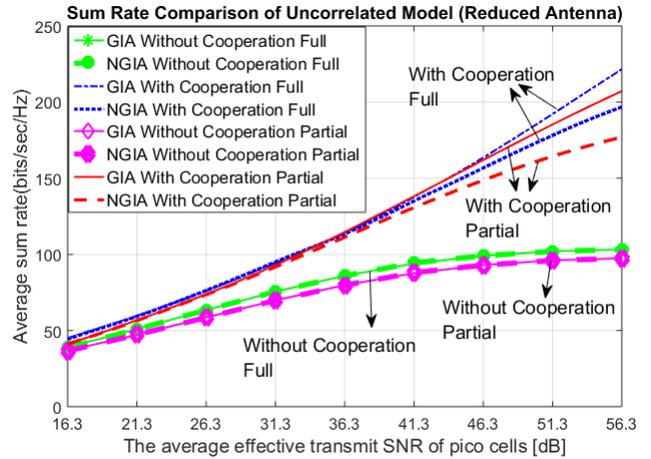

Fig. 8. Average sum rates vs. the effective transmit SNR of picocells in uncorrelated channel model (Reduced antenna).

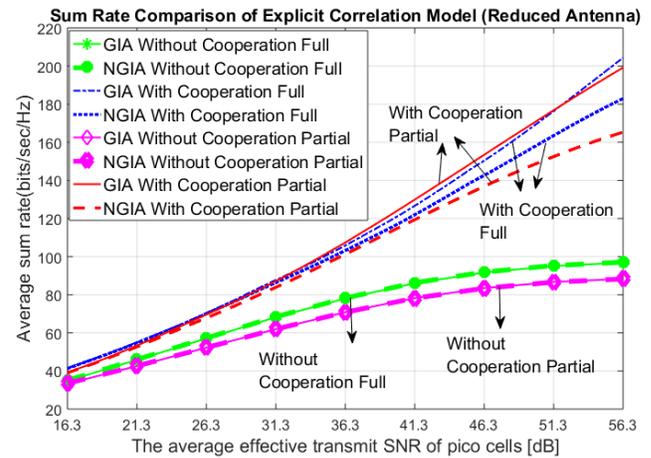

Fig. 9. Average sum rates vs. the effective transmit SNR of picocells in explicit correlation channel model (Reduced antenna).

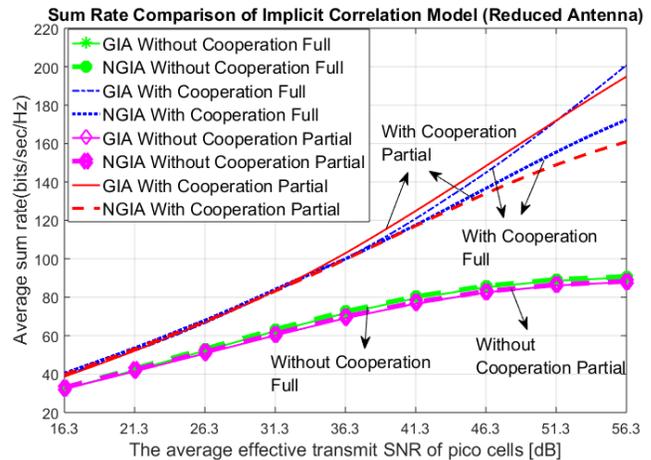

Fig. 10. Average sum rates vs. the effective transmit SNR of picocells in implicit correlation channel model (Reduced antenna).

In Fig. 8, Fig. 9 and Fig. 10, we separately consider uncorrelated channels, explicit correlation channels and 3GPP channels with perfect and imperfect CSI. We analyze the performance of various GIA designs as defined previously. The observations made in the previous section for the "sufficient" configuration also hold for the "insufficient" configuration in this section and will not be repeated here. It is remarkable that MMSE-type approaches work while IA does not work in this configuration with insufficient numbers of antennas. It is interesting to see in Fig. 9 and Fig. 10 that the sum rate of "with cooperation partial" is larger than "with cooperation full" at mid SNR values. This is because the number of antennas for each device is reduced. Then "full" data transmission scheme may cause more inter data stream interference than "partial" data transmission scheme, especially in correlated channels.

### 5.3 Complexity analysis

In this section, computational complexities of TSIA, GIA without cooperation, and GIA with cooperation are analyzed and compared. In the analysis below, we only consider the number of multiplications and the complexity of each multiplication is denoted by $O(1)$.

The complexity analysis of TSIA is made only for the iteration procedure in conventional IA (i.e. (34)). According to the assumptions mentioned above, computing $\mathbf{Z}_1$ has complexity $O(\sum_{j=2}^{L_1}(r_1 t_j m_j + r_1^2 m_j + r_1^2))$. The complexity of computing $\sum_{i=2}^{L_1} \mathbf{Z}_i$ is $O(\sum_{i=2}^{L_1}(r_i m_1 t_{hn} + r_i^2 m_1 + r_i^2 + \sum_{j=2, j \neq i}^{L_1}(r_i t_j m_j + r_i^2 m_j + r_i^2)))$ where $t_{hn} = t_1 - rank([h'_{L_1+1,1}, ..., h'_{L_1+n,1}]')$. Similarly, in the uplink scenario, complexity for $\overleftarrow{\mathbf{Z}}_1$ and $\sum_{j=2}^{L_1} \overleftarrow{\mathbf{Z}}_j$ are $O(\sum_{i=2}^{L_1}(t_{hn} r_i m_i + t_{hn}^2 m_i + t_{hn}^2))$ and $O(\sum_{j=2}^{L_1}(\sum_{i \neq j}^{L_1}(t_j r_i m_i + t_j^2 m_i + t_j^2)))$, respectively. As the complexity for eigenvalue decomposition (EVD) is $O(n^3)$ for a $n \times n$ matrix, $\sum_{i=1}^{L_1} \mathbf{G}_i = v_d(\mathbf{Z}_i)$ and $\sum_{j=1}^{L_1} \overleftarrow{\mathbf{G}}_j = v_d(\overleftarrow{\mathbf{Z}}_j)$ together has complexity $O(\sum_{i=1}^{L_1} r_i^3 + t_{hn}^3 + \sum_{j=2}^{L_1} t_j^3)$.

As for GIA algorithms, all $L$ cells are involved in the iteration procedure. Thus, computational complexity of GIA presented here is also for all cells. According to Strassen algorithm [26], matrix multiplication between two $n \times n$ matrices, and matrix inversion of a $n \times n$ matrix has complexity $O(n^{\log_2 7})$. Define $T = \sum_{i=1}^{L} t_i$. Then, for GIA without cooperation, complexities for *steps 1, 2, and 3* are

$O\left(\sum_{i=1}^{L}\left(t_i(2r_i m_i + m_i + m_i^2) + 2m_i r_i^2 + r_i^{\log_2 7}\right)\right)$,

$O\left(\sum_{i=1}^{L}(2 + t_i(2m_i^2 + r_i^2 + r_i T + r_i m_i + m_i + 1) + t_i^2(2m_i + r_i + T) + m_i(r_i^2 + T r_i + T^2)\right)$, and

$O\left(\sum_{i=1}^{L}(T^{\log_2 7} + T^2(r_i + m_i) + 2T r_i m_i)\right)$, respectively. For GIA with cooperation, complexities for *step 1, 2, and 3* are

$O\left(\sum_{i=1}^{L}\left(\sum_{j \in S}(T m_j + T^2 m_j) + T^2 r_i + T(r_i^2 + m_i^2 + m_i r_i) + r_i^{\log_2 7} + m_i r_i^2\right)\right)$,

$O\left(\sum_{i=1}^{L}\left(T^{\log_2 7} + t_i + 2 + \sum_{j \in S}(T(2m_j^2 + r_j^2 + m_j + 2r_j m_j) + T^2(3m_j + 2r_j) + r_j^2 m_j)\right)\right)$, and

$O\left(\sum_{i=1}^{L}\left(\sum_{j \in S}(T^2 m_j + m_j r_j T) + m_i T^2 + T^{\log_2 7} + T r_i m_i\right)\right)$, respectively.

Table II
AVERAGE ITERATION NUMERS (PERFECT CSI)

|  | Uncorrelated Model | Explicit Corr Model | Implicit Corr Model |
|---|---|---|---|
| TSIA | 1097 | 1262 | 1683 |
| GIA Without Partial | 68 | 66 | 63 |
| GIA Without Full | 47 | 47 | 49 |
| GIA With Partial | 299 | 327 | 186 |
| GIA With Full | 321 | 372 | 141 |

Table III
AVERAGE ITERATION NUMBER (IMPERFECT CSI)

|  | Uncorrelated Model | Explicit Corr Model | Implicit Corr Model |
|---|---|---|---|
| TSIA | 1025 | 1208 | 1757 |
| GIA Without Partial | 68 | 66 | 63 |
| GIA Without Full | 48 | 47 | 48 |
| GIA With Partial | 279 | 303 | 202 |
| GIA With Full | 355 | 331 | 188 |

For a specific analysis, we consider the MIMO configuration described in section 5.1. Average iteration numbers required to obtain the simulation results in section 5.1 for TSIA, GIA with and without cooperation are given in table II (perfect CSI) and table III (imperfect CSI).

It is observed from table II and III that average iteration numbers of GIAs are much less than that of TSIA for all the three channel models. Under perfect CSI condition, iteration numbers of TSIA are 16.1, 19.1 and 26.7 times of GIA without cooperation partial for uncorrelated, explicit correlated and implicit correlated channel models, respectively. As for the computational complexity of single iteration under current MIMO configuration network, TSIA has complexity $O(\sim 1.35 \times 10^3)$ which accounts for 5 cells in subsystem 1. On the other hand, computational complexities of single iteration for GIA without cooperation partial and full are $O(\sim 2.6 \times 10^5)$ and $O(\sim 3 \times 10^5)$, respectively, which account for all 10 cells.

Under perfect CSI condition, the corresponding total complexities (single complexity × iteration number) of GIA without cooperation partial are 5.5, 4.6, and 3.3 times of TSIA for uncorrelated, explicit correlated, and implicit correlated channel models, respectively. Similarly, the total complexities of GIA without cooperation full are 4.8, 4.1, and 3.2 times of TSIA for uncorrelated, explicit correlated, and implicit correlated channel models, respectively. Similar results are obtained under imperfect CSI condition. It should be noted that iteration numbers in computing total complexities are from table II and III. It can be expected that when GIAs only consider sub-system 1, the iteration numbers will be smaller than the ones presented in table II and III, thus the total computational complexities of GIA without cooperation will be reduced.

For GIA with cooperation algorithms, it is regarded as performance benchmark in our simulation due to its fully cooperated requirements. It also has highest complexity of single iteration which is $O(\sim 2.0 \times 10^6)$ and $O(\sim 2.9 \times 10^6)$ for partial and full scenarios, respectively.

## 6 Conclusion

In this work, we proposed a GIA MMSE transceiver design scheme for downlink MIMO HetNets and used a two-tier HetNet to compare the proposed GIA against the TSIA. We considered three practical channel models (the uncorrelated, explicit correlation channel models, and the 3GPP channel model) with both perfect and imperfect CSI conditions. In each channel model, we also considered two design scenarios (i.e. with cooperation and without cooperation among cells) and in each scenario, two data transmission schemes were taken into account. We investigated the influence of transmit power strength of pico BS on system performance. Complexity analysis and comparison between GIA and TSIA designs are also presented. The simulation results show that GIA has a broad application in different system configurations and conditions, and GIA outperforms TSIA under eight different system configurations and conditions, especially in the noise-limited regime. It is shown that GIA can significantly improve system performances in the "with cooperation" scenario.

It is remarkable that GIA is very flexible and can be applied to system configurations with insufficient antennas and still yields good system performances while TSIA is infeasible. In addition, GIA can flexibly utilize the equipped antennas to adapt to different data transmission schemes. It can transmit many more data streams than TSIA. It can also deal with multiple UEs per cell even though the numerical example shown here has only one UE per cell.

Although TSIA has slightly smaller complexity compared with GIA without cooperation algorithm, it can only deal with some specific network configurations. It has to mitigate the inter-cell interference in some pre-specified manners. Thus, it cannot adapt to network changes or channel variations conveniently. In contrast, GIA mitigates all interferences and system noise collectively and is able to deal with different network configurations, cooperation scenarios, data transmission schemes, channel variations, system changes, etc., in the same unified and systematic manner. These characteristics make GIA a very good choice for analyzing and/or designing various HetNets in practice.